\documentclass[conference]{IEEEtran}
\IEEEoverridecommandlockouts
\usepackage{cite}
\usepackage{amsmath,amssymb,amsfonts}
\usepackage{algorithmic}
\usepackage{graphicx}
\usepackage{textcomp}
\usepackage{xcolor}
\def\BibTeX{{\rm B\kern-.05em{\sc i\kern-.025em b}\kern-.08em
    T\kern-.1667em\lower.7ex\hbox{E}\kern-.125emX}}
\begin{document}

\title{Latent Confidence Alignment for LLM Self-Assessment}

\author{
\IEEEauthorblockN{Ting-Yu Chen}
\IEEEauthorblockA{\textit{Department of Information Management} \\
\textit{National Sun Yat-Sen University}\\
Kaohsiung, Taiwan \\
\phantom{x}brontotingyu@gmail.com}
\and
\IEEEauthorblockN{Tingting Yu}
\IEEEauthorblockA{\textit{Department of Information Management} \\
\textit{National Sun Yat-Sen University}\\
Kaohsiung, Taiwan \\
\phantom{x}tingyui0213@gmail.com}
\and
\IEEEauthorblockN{Pei-Cing Huang}
\IEEEauthorblockA{\textit{Department of Information Management} \\
\textit{National Sun Yat-Sen University}\\
Kaohsiung, Taiwan \\
\phantom{x}pcpeicing@gmail.com}
\and
\IEEEauthorblockN{Chan Hsu}
\IEEEauthorblockA{\textit{Department of Information Management} \\
\textit{National Sun Yat-Sen University}\\
Kaohsiung, Taiwan \\
\phantom{x}chanshsu@gmail.com}
\and
\IEEEauthorblockN{Ming-Yen Lin}
\IEEEauthorblockA{\textit{Kaohsiung Medical University Hospital} \\
\textit{Kaohsiung Medical University}\\
Kaohsiung, Taiwan \\
\phantom{x}mingyenlin3@gmail.com}
\and
\IEEEauthorblockN{Yihuang Kang}
\IEEEauthorblockA{\textit{Department of Information Management} \\
\textit{National Sun Yat-Sen University}\\
Kaohsiung, Taiwan \\
\phantom{x}ykang@mis.nsysu.edu.tw}
}
 
\maketitle

\begin{abstract}
Confidence calibration in large language models (LLMs) is commonly evaluated by comparing predicted confidence with observed accuracy. However, such approaches do not model item difficulty, making it difficult to interpret discrepancies and to determine whether model confidence reflects genuine self-assessment or is merely a byproduct of the response generation process. To address this, we adopt a Rasch model–based latent ability framework and a metacognitive perspective, and propose Latent Confidence Alignment Error (LCAE) to measure the consistency between model self-assessment and the latent error probability implied by model ability and item difficulty. We further incorporate item difficulty as an external signal with a reasoning mechanism. Experiments on a medical-domain dataset with 20 models show that the proposed approach improves self-assessment quality without affecting model ability, and reveals an association between reliability and inference cost.
\end{abstract}

\renewcommand{\IEEEkeywordsname}{Keywords}


\section{Introduction}
Large language models (LLMs) are increasingly being applied in high-stakes domains such as healthcare, law, and finance \cite{1}. Their reliability depends not only on performance on standardized benchmarks but also on their ability to assess their own uncertainty appropriately \cite{2}. As a result, model confidence and its calibration have become important aspects of reliability evaluation \cite{3}. Prior studies have shown that confidence assessment is influenced by task difficulty and the structure of the reasoning process \cite{4}.

Existing calibration metrics, such as Expected Calibration Error (ECE) \cite{5} and the Brier score \cite{6}, typically assess calibration by comparing predicted confidence with observed accuracy. However, these approaches rely on response outcomes from evaluation datasets and compute calibration error in an aggregated manner \cite{3}, making it difficult to characterize variation at the item level. Without explicitly modeling item difficulty \cite{7}, discrepancies between confidence and outcomes cannot be attributed to either inaccurate self-assessment or appropriate difficulty perception. In addition, model confidence is typically produced during response generation, rather than derived from an independent self-assessment process  \cite{2}, and therefore may not accurately reflect its judgment. These limitations reduce the interpretability of model self-assessment.

Confidence evaluation should be grounded in the conditional performance jointly determined by model ability and item difficulty, and used to examine whether model self-assessment is consistent with the expected performance implied by its position in the latent ability structure. Item Response Theory (IRT) provides a statistical framework that simultaneously models examinee ability and item difficulty \cite{8}. Recent studies have applied IRT to LLM evaluation and demonstrated its potential for estimating item difficulty \cite{7}. However, existing work remains largely limited to ability estimation and comparisons between models.

Based on the above background, this study adopts a metacognitive perspective \cite{9}, separating model self-assessment from post-reflection response outcomes as an independent analytical dimension, and further examines its relationship with model ability and inference cost \cite{10}, as well as the interrelationships among these factors in practical application scenarios. The key contributions of this study are summarized as follows:
\begin{itemize}
\item We propose an LLM evaluation framework that integrates IRT and metacognition, separating model ability comparison and self-assessment into two independent processes, and analyzing model performance across different evaluation dimensions.
\item We demonstrate, through the proposed Latent Confidence Alignment Error (LCAE), that models exhibit significant differences in the quality of their self-assessment even at comparable ability levels.
\item We systematically analyze the effects of external information and internal reflection mechanisms on model self-assessment, and reveal the key role of difficulty signals in adjusting model self-assessment.
\end{itemize}

The remainder of this paper is organized as follows. Section 2 reviews related work on LLM evaluation, including confidence estimation and calibration methods, the IRT measurement framework, and the role of reflection mechanisms in cognitive science. Section 3 presents the proposed method. Section 4 reports experimental results on a medical dataset to evaluate the applicability of the framework in high-stakes scenarios. Finally, Section 5 concludes the paper and summarizes the main findings.
\section{BACKGROUND}
In high-stakes domains, the outputs of LLMs may directly affect the quality of real-world decision-making \cite{1}. Existing evaluation methods typically rely on multiple benchmark datasets to compare model capabilities across tasks \cite{11}. To more finely characterize model capability, prior work has introduced item difficulty annotations within datasets \cite{12}. However, these annotations are often determined by human experts or task designers and may not reflect the difficulty as experienced by the model during response generation \cite{13}, \cite{14}.

IRT has been widely adopted in psychometrics and educational measurement \cite{8}, providing a probabilistic framework for jointly estimating latent ability and item parameters based on observed responses. Among its variants, the Rasch model (1PL) places examinee ability ($\theta$) and item difficulty ($b$) on the same latent scale and models the probability of a correct response as a logistic function of their difference \cite{15}. Compared to the 2PL and 3PL models, the Rasch model possesses the property of specific objectivity \cite{16}, meaning that comparisons between examinees are independent of the particular set of items, and comparisons between items are independent of the particular sample of examinees \cite{17}. Recent studies have applied the Rasch model to LLM benchmark evaluation, showing that the model can provide stable ability estimates even with a relatively small number of test items \cite{18}.

The aforementioned line of work provides a statistical foundation for modeling model capability based on observed responses, while self-assessment from the model’s own perspective constitutes another research direction. In cognitive science, the monitoring and regulation of one’s own cognitive states is referred to as metacognition [9], which centers on evaluating the uncertainty of one’s own judgments and adjusting subsequent decision-making and reasoning accordingly. Recent studies have extended this concept to LLMs by adopting metacognition-oriented prompting, guiding models to perform confidence estimation or reflection on their own answers \cite{19}, thereby producing estimates of the correctness of their judgments. Signal Detection Theory provides a cognitive framework for understanding such behaviors by distinguishing between two levels of judgment: Type-1 judgments correspond to task performance, whereas Type-2 judgments correspond to evaluations of the correctness of one’s own judgments \cite{20}. Empirical studies further show that when LLMs assess their confidence within the same context immediately after generating an answer, they may exhibit systematically inflated confidence, reflecting a tendency to assign higher confidence to their own generated responses \cite{2}. Therefore, task performance and self-assessment operate at different levels and should be evaluated separately.

Dual Process Theory provides a perspective for understanding different modes of reasoning \cite{21}, distinguishing between fast and intuitive System 1 and slower, more analytical System 2 \cite{22}. Prior studies have shown that LLMs can exhibit behaviors analogous to these two systems under different prompting strategies \cite{23}, and that the effectiveness of such strategies depends on both task characteristics and model architecture. In addition to internal reasoning processes, external information can also influence model reasoning and the expression of confidence. For example, incorporating external documents in retrieval-augmented generation has been shown to improve confidence calibration \cite{24}. Therefore, both adjusting reasoning strategies and incorporating external information may improve model self-assessment. 

This study integrates measurement theory from psychology with theoretical frameworks from cognitive science and proposes an evaluation approach for diagnosing LLM confidence based on latent model ability. It introduces item difficulty estimated from model response outcomes as external information, together with a mechanism that adaptively selects reasoning depth, to improve model self-assessment and analyze inference cost \cite{10}, thereby providing guidance for model selection that balances reliability and efficiency in high-stakes applications.

\section{IRT-Based Metacognitive Framework for LLM Evaluation}
We propose an evaluation framework based on a unified latent scale and, from a metacognitive perspective, analyze the consistency between model self-assessment and its expected performance to support model selection. As shown in Fig. 1, the framework consists of three stages: capability estimation, metacognitive decision making, and model selection.

In the first stage, each model answers the evaluation items. To enable unified analysis of model performance, the responses are converted into a binary response matrix, where correct answers are coded as 1 and incorrect answers as 0. Based on this response matrix, we apply the Rasch model to estimate model ability ($\theta$) and item difficulty ($b$). In the Rasch model, the probability that model  i correctly answers item q is defined as:

\begin{equation}
P_{iq} = P\left(X_{iq} = 1 \mid \theta_i, b_q\right) = \frac{1}{1 + e^{-(\theta_i - b_q)}}
\end{equation}
where $\theta_i$ represents the ability parameter of model $i$, and $b_q$ denotes the difficulty parameter of item $q$. When a model’s ability exceeds the difficulty of an item, the probability of answering the item correctly increases. 
\begin{figure*}[!t]
\centering
\includegraphics[width=0.95\textwidth]{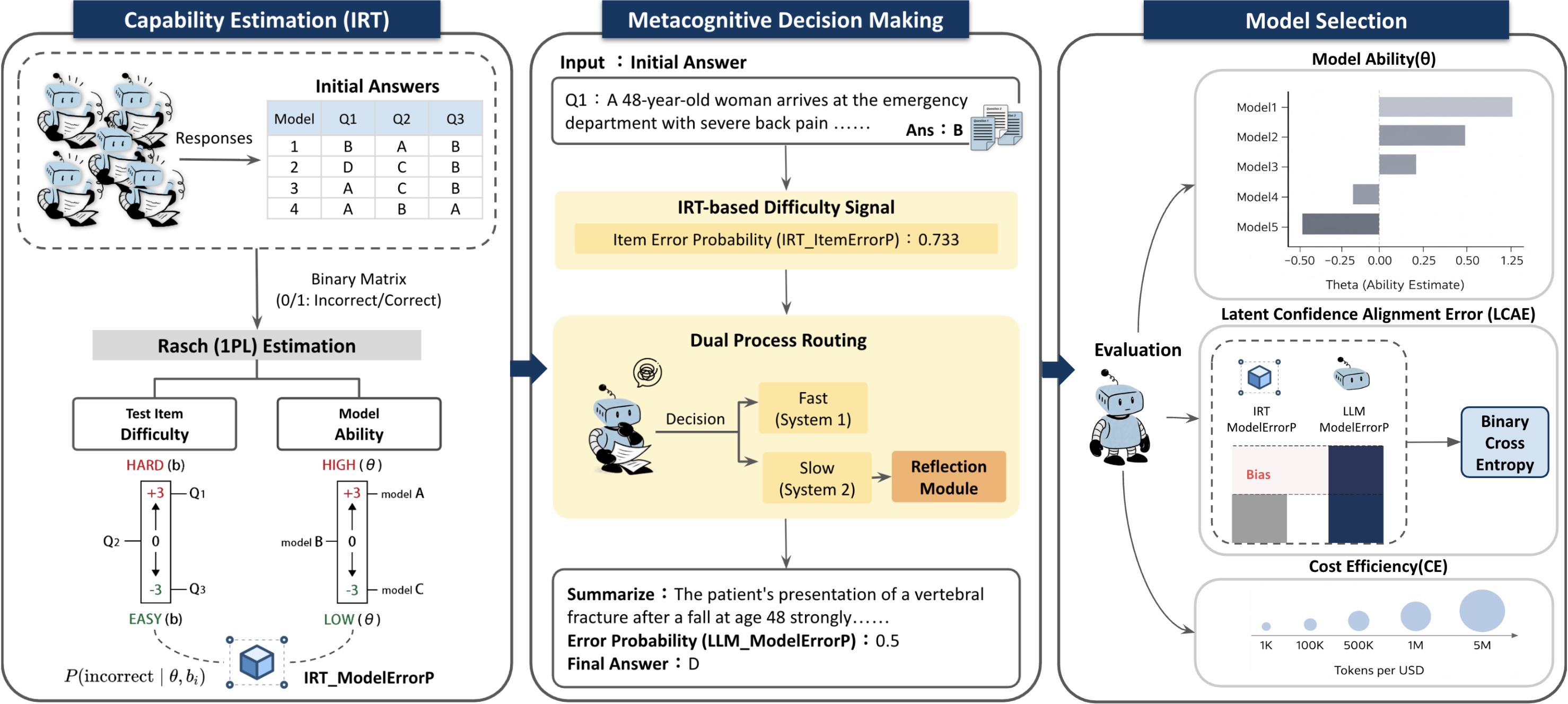}
\caption{Overview of the proposed three-stage LLM evaluation.}
\label{fig}
\end{figure*}

After estimating the ability and difficulty parameters, we further define two probability indicators of response error: {IRT Item Error Probability} (IRT\_ItemErrorP) and {IRT Model Error Probability} (IRT\_ModelErrorP). The former represents the average probability of an incorrect response to item $q$ across the set of evaluated models, capturing the combined effect of item difficulty and the distribution of abilities across the population of models. The latter measures the probability that model $i$ produces an incorrect answer on item $q$, derived from its estimated latent ability. Since these indicators are based on latent trait estimates rather than observed responses at any particular stage, they provide a stable latent reference for model evaluation in subsequent stages. The IRT\_ItemErrorP for item $q$ is defined as:
\begin{equation}
\mathrm{IRT\_ItemErrorP}_q = 1 - \frac{1}{I} \sum_{i=1}^{I} P_{iq}
\end{equation}
and the IRT\_ModelErrorP for model $i$ is defined as:
\begin{equation}
\mathrm{IRT\_ModelErrorP}_{iq}=1-P_{iq}
\end{equation}
where $I$ is the total number of models.

In the second stage, model self-assessment is treated as a stage independent of the initial response, and the model is guided through prompting to perform metacognitive decision making so that self-assessment can be compared with latent ability. The model is then given the question together with its initial answer from Stage~1. Based on this information, the model is prompted to output an error probability between 0 and 1 for its original answer, defined as the LLM Model Error Probability (LLM\_ModelErrorP), as its self-assessment. This estimate is then compared with the expected performance implied by the latent ability estimated in Stage~1. The model may then revise its original answer on the basis of its self-assessed error probability and produce a final answer, which represents its post-reflection performance.

Effective reasoning in complex tasks requires both sufficient information and an appropriate mode of thinking. To support this process, two components are introduced. First, the IRT-based Difficulty Signal (IDS) serves as external information and provides the model with $\mathrm{IRT\_ItemErrorP}$ to contextualize item difficulty during response self-assessment. Second, Dual-Process Routing (DPR) is introduced as a reasoning mechanism that enables the model to determine, based on its own judgment, whether to engage in reflection and to select between two reasoning modes: Fast (System~1) and Slow (System~2). When the model selects ``Fast,'' it generates an answer based on its existing knowledge and reasoning; when it selects ``Slow,'' it activates a reflection module to perform structured, step-by-step reflection on its prior answer. The module is a configurable component whose specific implementation needs to be defined according to task requirements. Through these components, the model can adjust its decisions during both self-assessment and reasoning.

In the final stage, we comprehensively evaluate the model. For task ability, we assess the model based on the final answers produced in the second stage. Because metacognition allows the model to revise its original answer through reflection and self-assessment, these final responses better reflect performance after the full decision-making process. Specifically, we re-estimate model ability using the Rasch model with item difficulty parameters fixed from Stage~1, ensuring that the estimates from different stages lie on the same latent scale.

To quantify potential discrepancies between performance and self-assessment, we define Latent Confidence Alignment Error (LCAE). This metric compares the model’s self-assessed error probability with a latent error probability derived from the IRT model, capturing the gap between the model’s internal estimate and an objective latent reference. For each model $i$, LCAE is defined as the binary cross-entropy (BCE) between LLM\_ModelErrorP and IRT\_ModelErrorP across all items, as follows~\cite{25}:

\begin{equation}
\mathrm{LCAE} 
= -\frac{1}{Q} \sum_{q=1}^{Q} 
\left[
e_{iq} \log\left(\hat{c}_{iq}\right) 
+ \left(1 - e_{iq}\right)\log\left(1 - \hat{c}_{iq}\right)
\right]
\end{equation}
where $Q$ denotes the total number of items. For a given model, $e_{iq}$ denotes the IRT-based latent error probability on item $q$, serving as a latent reference, and corresponds to $\mathrm{IRT\_ModelErrorP}$, while $\hat{c}_{iq}$ denotes the model’s self-assessed error probability on item $q$ and corresponds to $\mathrm{LLM\_ModelErrorP}$ obtained in the metacognitive stage. A lower $\mathrm{LCAE}$ value indicates that the model’s estimated error probability is more consistent with the latent error probability, meaning that the model’s self-assessment more closely reflects its expected performance under the latent ability structure.

In cost evaluation, we introduce Cost Efficiency (CE) to enable relative cost comparison across models. CE measures tokens generated per unit cost. To ensure consistency, we adopt a simplified formulation given variations in input and output token usage~\cite{26}.
\begin{equation}
\bar{C} = \frac{P_{in} + P_{out}}{2}
\end{equation}
where $P_{\mathrm{in}}$ and $P_{\mathrm{out}}$ denote the input and output prices per million tokens, respectively. The CE is then defined as the number of tokens generated per USD:
\begin{equation}
CE = \frac{10^6}{C}
\end{equation}
where a higher CE indicates that more tokens can be generated per unit cost. 

Through the ability parameter, $\mathrm{LCAE}$, and $\mathrm{CE}$, we analyze models from multiple perspectives, thereby supporting comparison and selection across different models.

\section{EXPERIMENTS AND DISCUSSION}
To evaluate the proposed framework in a high-stakes setting, we used MedXpertQA as the primary benchmark dataset~\cite{27}. The dataset covers diverse clinical tasks and has been validated by experts to ensure data quality. This study is restricted to text-based questions, and 100 items were randomly sampled using a fixed random seed of 42 to ensure reproducibility.

In terms of model configuration, this study selected 20 LLMs, covering mainstream model families widely evaluated in the medical domain, including Claude, DeepSeek, Gemini, Llama, GPT, and Qwen~\cite{28},\cite{29},\cite{30}, with representative versions included. For all models, the temperature was set to 0 to reduce randomness and ensure deterministic outputs. Because prompting conditions can substantially affect model performance~\cite{31}, each model–prompt combination was treated as a separate examinee to increase the effective sample size. Models were paired with five prompting strategies, namely Standard Prompting, Knowledge Prompting, Chain-of-Thought (CoT), Self-Ask, and Tree-of-Thought (ToT)~\cite{32}, resulting in 100 examinees for stable Rasch parameter estimation.
\begin{table}[!htbp]
\refstepcounter{table}\label{tab:cb_pairwise}
\centering
\footnotesize
\text{TABLE \thetable.}\par
\textsc{Pairwise comparisons of lcae across conditions.}\par
\vspace{4pt}
\renewcommand{\arraystretch}{1.05}
\begin{tabular}{lccc}
\multicolumn{4}{l}{\textbf{Friedman:} $\chi^2(3)=35.15$, $p<.001$} \\
\noalign{\hrule height 1pt}
\textbf{Pairwise Comparison} & \textbf{p-value} & \textbf{Adj. p-value} & \textbf{diff} \\
\noalign{\hrule height 0.6pt}
Baseline - DPR    & 0.330          & 0.330           & +0.270 \\
Baseline - IDS     & \textbf{0.001}  & \textbf{0.001}   & +0.672 \\
Baseline - IDS+DPR & \textbf{0.001}  & \textbf{0.001}   & \textbf{+0.829} \\
DPR - IDS          & \textbf{0.012} & \textbf{0.036} & +0.402 \\
DPR - IDS+DPR      & \textbf{0.001}  & \textbf{0.001}   & +0.560 \\
IDS - IDS+DPR       & 0.097          & 0.195           & +0.157 \\
\noalign{\hrule height 1pt}
\end{tabular}
\end{table}

\begin{table}[!htbp]
\refstepcounter{table}\label{tab:ability_pairwise}
\centering
\footnotesize
\text{TABLE \thetable.}\par
\textsc{Pairwise comparisons of ability across conditions.}\par
\vspace{4pt}
\renewcommand{\arraystretch}{1.05}
\begin{tabular}{lccc}
\multicolumn{4}{l}{\textbf{Friedman:} $\chi^2(3)=0.52$, $p>0.05$} \\
\noalign{\hrule height 1pt}
\textbf{Pairwise Comparison} & \textbf{p-value} & \textbf{Adj. p-value} & \textbf{diff} \\
\noalign{\hrule height 0.6pt}
Baseline - DPR    & 0.928 & 1.00 & -0.000 \\
Baseline - IDS     & 0.674 & 1.00 & +0.007 \\
Baseline - IDS+DPR & 0.615 & 1.00 & -0.022 \\
DPR - IDS          & 0.675 & 1.00 & +0.008 \\
DPR - IDS+DPR      & 0.520 & 1.00 & -0.021 \\
IDS - IDS+DPR       & 0.520 & 1.00 & -0.029 \\
\noalign{\hrule height 1pt}
\end{tabular}
\end{table}
Following the Rasch modeling procedure, item fit was evaluated using infit and outfit mean square statistics. Among the 100 items, 90 fell within the acceptable range (0.5--1.5), while the remaining items slightly exceeded this range but did not reach the threshold for severe misfit ($>2.0$). These deviations may reflect inherent variability in model responses, possibly 
arising from the stochastic nature of LLM generation~\cite{33}. Therefore, all items were retained to preserve the completeness and stability of the evaluation set.

In the metacognitive decision-making stage, to control experimental cost and maintain analytical consistency, this study focuses on the subset of examinees under the Standard Prompting condition in Stage~1, and all subsequent experiments are conducted under a single reasoning setting to avoid variability introduced by different prompting strategies, allowing comparisons to focus on model decision behavior across mechanism conditions. Under this setting, four conditions are designed for comparison: baseline, IDS, DPR, and IDS with DPR. Under the baseline condition, the model performs self-assessment based only on the question and its initial response. Under the IDS condition, IRT\_ItemErrorP is provided as an external difficulty signal to reflect item difficulty. Under the DPR condition, a reflection mechanism is introduced, and the NAP (Narrow, Analyze, Probe) component of the SNAPPS framework~\cite{34} is adopted as the design; specifically, the model is guided through three steps: \textit{Narrow} focuses on plausible candidate diagnoses, \textit{Analyze} evaluates the consistency between diagnoses and clinical features, and \textit{Probe} identifies key information gaps and sources of uncertainty in the reasoning process. Under the IDS with DPR condition, the model incorporates both external difficulty information and the reflection mechanism to evaluate the combined effect of the two mechanisms.

\begin{figure*}[!t]
\centering
\includegraphics[width=0.88\textwidth]{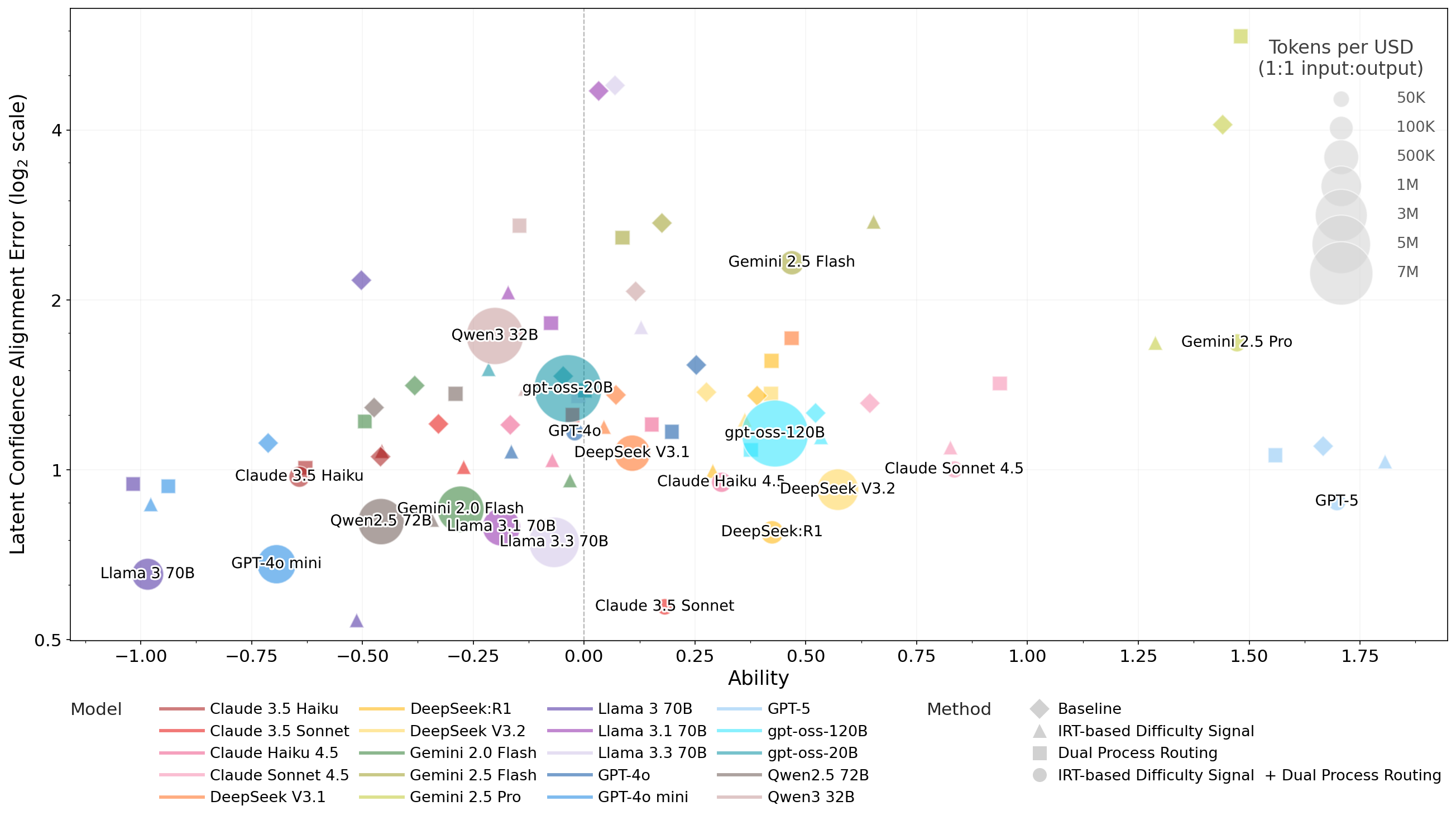}
\caption{Multi-dimensional performance of 20 LLMs across ability, LCAE, and CE under different conditions.}
\label{fig2}
\end{figure*}

\begin{table}[!htbp]
\refstepcounter{table}\label{tab:top7_models}
\centering
\footnotesize
\text{TABLE \thetable.}\par
\textsc{Top 7 LLMs ranked by ability across conditions.}\par
\vspace{4pt}
\renewcommand{\arraystretch}{1.05}
\setlength{\tabcolsep}{6pt}
\begin{tabular}{lccc}
\noalign{\hrule height 1pt}
\textbf{Method} & \textbf{Ability}$\uparrow$ & \textbf{LCAE [CI95\%]}$\downarrow$ & \textbf{CE}$\uparrow$ \\
\noalign{\hrule height 0.6pt}

Claude Sonnet 4.5 & 0.631 & 1.308 [1.177, 1.436] & \\
+DPR              & 0.995 & 1.423 [1.217, 1.693] & \raisebox{-1.8ex}[0pt][0pt]{111K} \\
+IDS              & 0.814 & \textbf{1.021 [0.907, 1.147]} & \\
+DPR+IDS          & 0.814 & 1.081 [0.962, 1.207] & \\
\noalign{\hrule height 0.6pt}

DeepSeek:R1       & 0.376 & 1.440 [1.277, 1.610] & \\
+DPR              & 0.396 & 1.528 [1.326, 1.741] & \raisebox{-1.8ex}[0pt][0pt]{625K} \\
+IDS              & 0.309 & 0.947 [0.836, 1.084] & \\
+DPR+IDS          & 0.418 & \textbf{0.866 [0.753, 0.982]} & \\
\noalign{\hrule height 0.6pt}

DeepSeek V3.2     & 0.328 & 1.331 [1.167, 1.487] & \\
+DPR              & 0.440 & 1.398 [1.217, 1.633] & \raisebox{-1.8ex}[0pt][0pt]{3.08M} \\
+IDS              & 0.360 & 1.215 [1.042, 1.405] & \\
+DPR+IDS          & 0.630 & \textbf{0.965 [0.825, 1.121]} & \\
\noalign{\hrule height 0.6pt}

Gemini 2.5 Flash  & 0.220 & 2.725 [2.211, 3.314] & \\
+DPR              & 0.108 & 2.635 [2.135, 3.211] & \raisebox{-1.8ex}[0pt][0pt]{714K} \\
+IDS              & 0.714 & 2.673 [2.128, 3.353] & \\
+DPR+IDS          & 0.420 & \textbf{2.233 [1.825, 2.692]} & \\
\noalign{\hrule height 0.6pt}

Gemini 2.5 Pro    & 1.445 & 4.129 [3.455, 4.879] & \\
+DPR              & 1.464 & 5.870 [4.858, 6.933] & \raisebox{-1.8ex}[0pt][0pt]{178K} \\
+IDS              & 1.347 & \textbf{1.588 [1.382, 1.815]} & \\
+DPR+IDS          & 1.483 & 1.717 [1.432, 1.996] & \\
\noalign{\hrule height 0.6pt}

GPT-5             & \textbf{1.677} & 1.058 [0.938, 1.176] & \\
+DPR              & \textbf{1.619} & 1.038 [0.927, 1.155] & \raisebox{-1.8ex}[0pt][0pt]{178K} \\
+IDS              & \textbf{1.755} & 0.999 [0.889, 1.120] & \\
+DPR+IDS          & \textbf{1.639} & \textbf{0.941 [0.840, 1.041]} & \\
\noalign{\hrule height 0.6pt}

gpt-oss-120B      & 0.483 & 1.162 [1.048, 1.283] & \\
+DPR              & 0.377 & 1.101 [0.995, 1.213] & \raisebox{-1.8ex}[0pt][0pt]{\textbf{8.73M}} \\
+IDS              & 0.482 & 1.112 [0.992, 1.244] & \\
+DPR+IDS          & 0.462 & \textbf{1.061 [0.962, 1.163]} & \\
\noalign{\hrule height 1pt}

\end{tabular}
\end{table}

To compare the four conditions, we conducted a Friedman test with pairwise Wilcoxon signed-rank tests and Holm--Bonferroni correction ($\alpha = 0.05$). LCAE results (Table~I) show a significant overall effect ($\chi^2(3) = 35.15$, $p < 0.001$). IDS+DPR significantly reduces LCAE compared to the baseline, whereas DPR alone shows no significant effect. Ability results (Table~II) show no significant differences across conditions, indicating no observable change in model ability.

We next examine model performance under different conditions. Table~III presents the results for the top seven models ranked by ability across four conditions (baseline, DPR, IDS, and DPR with IDS). Each condition was repeated three times, and the mean LCAE with 95\% confidence intervals is reported. Notably, even among models ranked at the top in terms of ability, LCAE under the baseline condition still exhibits substantial variation, indicating that higher ability does not necessarily correspond to consistent self-assessment quality.

Introducing IDS consistently reduces LCAE across models relative to their respective baselines, and in most cases, combining DPR with IDS leads to further reductions. In contrast, DPR alone does not yield consistent improvements; for some models, LCAE even increases. For example, Gemini~2.5~Pro~\cite{38} shows an increase from 4.129 under the baseline to 5.870 under DPR, suggesting that reflection alone may not reliably improve self-assessment alignment without external difficulty information. The magnitude of improvement varies across models. GPT-5~\cite{39} exhibits relatively low LCAE at baseline (1.058) and shows only limited changes under additional mechanisms, indicating stable self-assessment alignment. In contrast, DeepSeek~R1~\cite{40} and DeepSeek~V3.2~\cite{41} achieve their lowest LCAE under the combined DPR and IDS condition, suggesting that the integration of reflection and difficulty information can provide additional benefits for some models.

Fig.~2 provides an overall view of all twenty models across three dimensions: ability, LCAE, and cost efficiency (CE). Since ability is a latent parameter estimated based on the full set of evaluated models, it is more appropriately interpreted in terms of its relative position within the overall distribution rather than from local rankings alone. As shown in the figure, ability and LCAE do not exhibit a monotonic relationship. In addition, gpt-oss-120B~\cite{42} maintains relatively low LCAE (1.061) under the combined DPR and IDS condition while achieving the highest CE (8.73M), suggesting that favorable self-assessment alignment can be attained alongside high inference cost efficiency.

This study shows that, in high-stakes medical settings, models differ not only in their capabilities but also in the alignment between their self-assessments and the error probabilities implied by the latent ability structure. Results indicate that incorporating item difficulty information significantly reduces LCAE, whereas reflection alone has a limited effect, suggesting that model self-assessment is more sensitive to external difficulty signals. Furthermore, joint analysis of ability, LCAE, and cost efficiency reveals that these dimensions do not exhibit consistent relationships, highlighting the multi-dimensional nature of model performance.

The framework still has several limitations. This study was validated only in the medical domain, and its applicability to other tasks remains to be explored. In addition, the limited experimental scale may affect the stability of the results. In the IDS mechanism, the introduced signal is derived from the same set of evaluated models, which may introduce bias, and its generalizability to unseen models requires further validation. The cost evaluation adopts a simplified estimation method, whose suitability depends on practical application needs. Nonetheless, the framework jointly analyzes model ability, self-assessment alignment, and inference cost, and improves alignment without requiring model retraining, thereby enhancing model selection quality.

\section{CONCLUSION}
This study proposes an LLM evaluation framework that integrates IRT with a metacognitive perspective, separating model ability estimation and self-assessment, and analyzing them on a shared latent scale. It further introduces Latent Confidence Alignment Error (LCAE) to measure the consistency between model self-assessment and latent error probability. The results show that, even at similar ability levels, models exhibit significant differences in self-assessment alignment, indicating that higher ability does not necessarily correspond to better self-assessment. External difficulty signals are the primary factor influencing model self-assessment; reflection alone does not yield a significant effect, but their combination further improves overall alignment. Overall, the framework enables joint analysis of model ability and self-assessment consistency on a shared latent scale, reveals differences across evaluation dimensions and their relationship with cost efficiency, and provides a multi-dimensional basis for model selection.
\section*{Acknowledgment}

The study is supported by the National Health Research Institutes, Taiwan (grant number: NHRI-EX115-11208PI).

\end{document}